\documentclass[aps]{revtex4}
\usepackage[dvips]{epsfig}
\usepackage{amsmath,amssymb}

\parskip=\medskipamount

\newcommand{\eq}[1]{(\ref{#1})}
\newcommand{\fig}[1]{Fig.\ref{#1}}

\newcommand{\be}{\begin{equation}}
\newcommand{\ee}{\end{equation}}

\newcommand{\barr}{\begin{array}}
\newcommand{\earr}{\end{array}}

\newcommand{\beqn}{\begin{eqnarray}}
\newcommand{\eeqn}{\end{eqnarray}}

\newcommand{\bs}{\begin{subequations}}
\newcommand{\es}{\end{subequations}}

\newcommand{\bw}{\begin{widetext}}
\newcommand{\ew}{\end{widetext}}

\newcommand\disp{\displaystyle}

\begin{document}

\title{Necklace--Cloverleaf Transition in Associating RNA--like Diblock
Copolymers}

\author{M.V.Tamm$^{1,2}$, S.K.Nechaev$^{1,3}$}

\affiliation{$^1$LPTMS, Universit\'e Paris Sud, 91405 Orsay Cedex,
France \\ $^2$Physics Department, Moscow State University 119992
Moscow, Russia \\ $^3$Landau Institute for Theoretical Physics,
117334 Moscow, Russia}

\date{June 20, 2006}

\begin{abstract}
We consider a ${\rm A}_m{\rm B}_n$ diblock copolymer, whose links are capable of
forming local reversible bonds with each other. We assume that the resulting
structure of the bonds is RNA--like, i.e. topologically isomorphic to a tree. We
show that, depending on the relative strengths of A--A, A--B and B--B contacts, such
a polymer can be in one of two different states. Namely, if a self--association is
preferable (i.e., A--A and B--B bonds are comparatively stronger than A--B contacts)
then the polymer forms a typical randomly branched cloverleaf structure. On the
contrary, if alternating association is preferable (i.e. A--B bonds are stronger
than A--A and B--B contacts) then the polymer tends to form a generally linear
necklace structure (with, probably, some rear side branches and loops, which do not
influence the overall characteristics of the chain). The transition between
cloverleaf and necklace states is studied in details and it is shown that it is a
2nd order phase transition.
\end{abstract}

\maketitle

\section{Introduction}
\label{sect:1}

It would not be a strong exaggeration to say that since the very beginning the
statistical physics of macromolecules was driven mainly by biological motivations.
The problems of phase rearrangements in biopolymers, such as proteins, DNA and RNA
molecules, have stimulated an intensive development of new theoretical approaches
aiming to describe and to predict the structural, conformational and functional
changes in biopolymers under variation of external conditions (see, for example
\cite{edw65,lif68,de_gennes68}).

A very important biological role has a wide class of so-called "associating"
polymers. The associating polymers, besides the strong covalent interactions
responsible for the very chain--like structure of a macromolecule with frozen
sequence ("primary structure") of monomer units in the chain, are capable of forming
additional weaker reversible temperature--dependent (i.e. "thermoreversible") bonds
between different monomer units. Many biologically important macromolecules, like
proteins and nucleic acids, belong to the class of associating polymers. It is known
that just the presence of thermoreversible bonds is crucial for formation of various
equilibrium spatial conformations of such macromolecules (so-called, ternary
structures), and, hence, affects biological activity of these macromolecules
\cite{ternary}.

The physics of associating polymers is typical for statistical mechanics of complex
systems and consists in an interplay between several entropic and energetic factors.
In the particular case of associating polymers one can pick out the following three
major factors, which, together with the primary structure of the chain, give rise to
all the variety of possible thermodynamic states in these systems:

\begin{itemize}
\item[--] the energy gain due to the formation of thermoreversible contacts;
\item[--] the combinatoric entropy due to the choice of which particular monomers
(among those able to participate in bonds formation) actually do create bonds;
\item[--] the loss of conformational entropy of the polymer chain due to bonds
formation, and in particular, the entropic penalty of loop creation between two
bonded monomers.
\end{itemize}

Among a variety of macromolecular systems with thermoreversible interactions we pay
a special attention to a class of so-called "RNA--like" polymers. These polymers are
distinguished from other biologically active associating polymers, such as, for
instance, proteins, by a capability of forming only "cloverleaf--like" (or
"cactus--like") secondary structures. Indeed, the formation of a thermoreversible
contact between two distant bonds in a RNA molecule (or in a single--stranded DNA)
imposes a nonlocal constraint on a number of other possible bonds: all bonds in a
RNA chain are known to be arranged in a way to allow only hierarchical cactus--like
folded conformations topologically isomorphic to a tree. The pairs of bonds, which
do not obey such a structure are called "pseudoknots" and are forbidden for
RNA--type molecules. A quantitative definition of conformations allowed and
forbidden in RNA--like polymers is described in the Section \ref{sect:2}. Note, that
for proteins such a constraint is relaxed.

There is plenty of works concerning the statistical properties of associating
polymers without any constraints on the topology of thermoreversible contacts
\cite{LGK76,assoc}. For such systems one can construct a satisfactory mean--field
theory \cite{LGK76}. However RNA--like polymers are much less investigated.

\begin{figure}[ht]
\epsfig{file=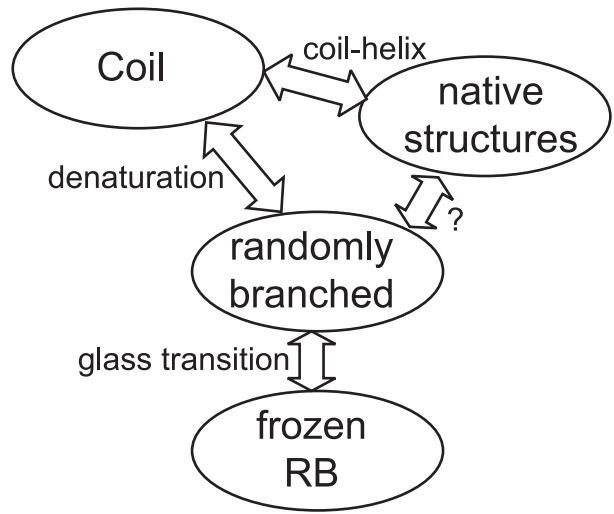,width=8cm} \caption{Schematic picture of possible
thermodynamic states of RNA--like heteropolymers.} \label{fig:1}
\end{figure}

It is known that depending on the temperature (which comes into play via the
temperature dependence of the statistical weights of the bonds) and on the primary
structure of the chain, the RNA--like polymer may be found in the following four
distinct states---see \fig{fig:1}. First, at high enough temperature any RNA--like
molecule forms a usual Gaussian or non-self--intersecting (depending on the quality
of the solvent) coil with no, or almost no reversible bonds formed. Second, at lower
temperature it may form a highly branched "genuine cactus--like" state topologically
isomorphic to a randomly branched tree. Third, there is a strong evidence that at
low enough temperatures the RNA molecules with highly irregular primary structure
tend to form glassy states. Finally, in the case of some special primary structures,
there exists an energy gap between the ground (so-called "native") energy state of
the RNA and all the other states. The RNA can, therefore, in this case form a highly
deterministic secondary structure (as compared to annealed random and quenched
random states in the previous two cases).

In particular, if the primary structure of a RNA consists of two (or several)
complimentary parts, it tends to form a double--folded (or, in case of several parts
-- a cross--like, as in typical tRNA) structure. Such a structure, contrary to a
randomly branched one, is highly deterministic --- one can predict easily which
particular monomers will aggregate with each other --- and has a different
asymptotic (for large chain lengths $N$) mean--square end-to-end distance, $R$. For
example, if one neglects the volume interactions (i.e. the non-saturating
interactions between monomers which are far from each other along the chain),
$R_{\rm nc} \sim N^{1/2}$ for necklace (or "double folded") conformation, rather
than $R_{\rm rb} \sim N^{1/4}$ in the randomly branching case.

Let us give a sketch of what is known about the transitions between the four states
depicted in the \fig{fig:1}. First of all, the transition between the
single--stranded and the double--stranded RNA (or DNA) commonly known as a
helix--coil transition, which is a typical representative of the transition between
the coil and the deterministic ground state, is rather well understood
\cite{Gros_Khokh}, the modern developments in this field can be found in
\cite{peliti}. It is known that the particular form of loop penalties plays crucial
role in whether it is a genuine phase transition or just a collective one. The
contribution to the polymer partition function emerging from the loops is usually
supposed to have the scaling form $N^{\alpha}$ (where $N$ is the loop length)---see,
for example, \cite{Gros_Khokh}. If $0\le \alpha<1$ ($\alpha=0$ corresponds to the
absence of any loop contribution) the passage from a coil to a helix is not a phase
transition, but a cooperative phenomenon with a more or less (depending on the value
of $\alpha$ and on the cooperativity of bond formation) sharp crossover. On the
contrary, if $\alpha \geq 1$ there is a genuine phase transition. The situation
concerning randomly branched RNA molecules is more complicated. Despite the
denaturation of a cactus--like RNA has been investigated for several decades,
starting from classical works \cite{de_gennes68,erukh78}, it was shown only recently
\cite{mueller} that the character of transition in RNA resembles in some aspects the
coil--helix transition in DNA. Namely, for large loop penalties ($\alpha \geq 2$ for
RNA) one has a genuine phase transition, while for smaller loop penalties there is
just a crossover behavior.

Note, that in both aforementioned cases the heteropolymer nature of real RNA
structure played a minor role in the transition under discussion: one could have
considered all monomers as identical and still have a proper insight into what goes
on. On the contrary, in the problem of glass transition in RNA--like polymers, the
frozen heteropolymer structure of a chain plays a crucial role: the presence of
frustrations in the system is itself a peculiarity of heterogenous systems
\cite{bundhwa}. There is, hence, a barest necessity for a statistical theory of {\it
heteropolymers} forming RNA--like structures. The investigation of some
thermodynamic properties of random heteropolymers is addressed in a few recent
theoretical papers \cite{bundhwa,mezmul,orland,weise}. On the other hand, the
theoretical results are usually obtained not for "real" heteropolymer with finite
number of link types, but for polymers with frozen Gaussian distribution of
association constants, sometimes even excluding the loop factor. It is believed
(see, for example, \cite{bundhwa}) that the loop penalties are less significant for
understanding the physical properties of RNA--like molecules at low temperature
(i.e., close to the glass transition) when the typical sizes of loops become rather
small.

We would like to stress that for the best of our knowledge, the theoretical
consideration of the possible transition between regular (for example,
double--stranded) and randomly branched cactus--like states of RNA is absent. It
seems to be instructive, therefore, to consider a model of a RNA--like polymer which
shares the peculiarity of a chain non-uniformity in a way, which can give rise to
the formation of linear double--stranded structure, with an advantage of being
exactly solvable. Presenting here such a model, we believe that besides of solving
specific model of the transition between double--stranded and randomly--branched
states of RNA, we also provide a better insight into the structure of possible
states of heteropolymer RNAs, thus contributing in the construction of the general
theory of heteropolymers forming RNA--like structures.

Specifically, we consider in this paper a ${\rm A}_m{\rm B}_n\equiv \overbrace{\rm
AA...A}^{m\; \rm times}\; \overbrace{\rm BB...B}^{n\; \rm times}$ diblock copolymer,
whose links can form local reversible bonds with each other. We show that, depending
on the relative strengths of A--A, A--B and B--B bonds, such a polymer can be in
one of two different states. Namely, if a self--association is preferable (i.e.,
A--A and B--B bonds are comparatively stronger than A--B ones) then the polymer
forms a typical randomly branched cloverleaf structure. On the contrary, if
alternating association is preferable (i.e. A--B bonds are stronger than A--A and
B--B contacts) then the polymer tends to form a generally linear necklace structure
somewhat reminiscent of a double--stranded DNA (with, probably, some rear side
branches and loops, which, however do not influence the overall characteristics of
the chain). In this paper we consider the transition between these two states
(cloverleaf and necklace) in detail, and show that it turns out to be a 2nd order
 phase transition.

The paper is organized as follows. In the Section \ref{sect:2} we discuss the model
under consideration, and derive the basic equations for the partition function. In
Section \ref{sect:3}, which plays the central role in the paper, we solve these
equations in the approximation of no loop penalties, and study the case of large
equal lengths of blocks, $m=n\gg 1$. We show that a necklace--cloverleaf transition
is a 2nd order phase transition, compute the corresponding phase diagram and discuss
the behavior of the free energy near the transition point. In the Section
\ref{sect:4} we solve the basic equations with "ideal" loop weights ($\alpha=3/2$)
and show that the presence of these loop factors does not have a strong influence on
the phase transition. Finally, in the Section \ref{sect:5} we summarize the results
and briefly discuss the perspectives.

\section{The model}
\label{sect:2}

\begin{figure}[ht]
\epsfig{file=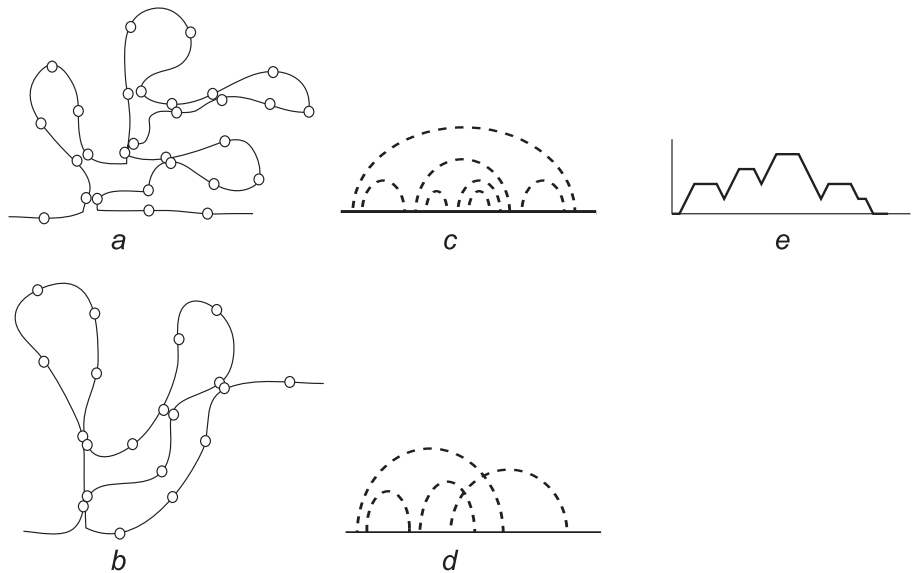,width=12cm} \caption{(a,b): Schematic picture of allowed
(a) cactus--like and prohibited (b) pseudoknot configurations of the bonds; (c,d):
Arc diagrams corresponding to configurations (a) and (b), respectively (note the
intersection of arcs in (d); (e): The height diagram corresponding to the bond
configuration (a).} \label{fig:2}
\end{figure}

Consider a ${\rm A}_n {\rm B}_m$ diblock copolymer whose monomers A and B are
capable of forming thermoreversible bonds of all possible types A--A, B--B, and
A--B. Let the statistical weights of bonds be equal $v_1$, $v_2$, and $u$,
respectively. Having the applications to RNA molecules in mind, assume also that the
structures formed by these thermoreversible bonds are always of a tree--like type,
as shown in \fig{fig:2}a. It means that we restrict ourselves to the situation in
which the chain conformations with "pseudoknots" shown in \fig{fig:2}b are
prohibited.

This definition can be formalized as follows. Take a chain ${\cal C}$ and enumerate
its links: $1, 2,..., n$. Consider two different thermoreversible bonds $b(i_1,i_2)$
and $b(j_1,j_2)$, where $b(i_1,i_2)$ corresponds to a contact between two monomers
$i_1$ and $i_2$ (where $1\le i_1<i_2\le n$), and $b(j_1,j_2)$ is formed by monomers
$j_1$ and $j_2$ (where $1\le j_1<j_2\le n$). We say that the chain ${\cal C}$ forms
a RNA--like structure if and only if {\it for any two different} bonds $b(i_1,i_2)$
and $b(j_1,j_2)$ the following inequality holds:
\be
(j_1-i_1)(j_2-i_1)(j_1-i_2)(j_2-i_2)>0
\label{eq:RNA}
\ee
To make the difference between allowed and not allowed structures geometrically more
transparent, redraw the structures shown in \fig{fig:2}a and \fig{fig:2}b in the
following way. Represent a polymer under consideration as a straight line with
active monomers situated along it in the natural order, and depict the
thermoreversible bonds by dashed arcs connecting the corresponding monomers. Now,
the absence of pseudoknots simply means the that the arcs do not intersect --- see
\fig{fig:2}c,d.

Moreover, we assume for simplicity, that except pseudoknots, all
other bond configurations are allowed. This means, in particular,
that we do not require any minimal loop length, as well as we do
not yet take into account the cooperativity effect (the fact that
if two links are connected with each other, then the two adjacent
links have larger probability to be also connected). These
assumptions are known to be false for real RNA molecules (for
example, there are no loops shorter than 3 monomers in RNA chains
\cite{mueller}). However, one can speculate that --- see, for
example, \cite{GGS1} --- if the links of the chain are considered
as renormalized quasi--monomers consisting of several "bare"
monomer units, the made assumptions seem to be plausible.

Let us also mention another possible representation of the
secondary structures (configurations of bonds) allowed in the
system under discussion. Consider again the arc diagram
representation. Let then unconnected monomers, the monomers at
which any arc starts and those where it ends correspond to vectors
$(1,0)$, $(1,1)$, and $(1,-1)$, respectively. Then any bond
configuration is mapped to some sequence of vectors which can be
considered as a random walk in (1+1)D (see \fig{fig:2}e) situated
in the upper half--plane. The walk begins and ends on the
$x$--axis; the end points are separated by the distance $N$ ---
the total length of the polymer. Such a representation is often
called the "height diagram" of the secondary structure. Note, that
the height of the point in the phase diagram is exactly equal to
the number of arcs going above the corresponding point on the arc
diagram, i.e. to the number of bonds one have to break to reach
the corresponding monomer from the starting point of the chain.
The critical exponent $\gamma$ connecting the mean height of such
a diagram with the length of the polymer ($\langle h\rangle \sim
N^{\gamma}$, $0\leqslant \gamma \leqslant 1$), called roughness
exponent, is an important characteristic of the state of the
system. In the randomly branched homopolymer state $\gamma_{\rm
br} = 1/2$, while in the glassy state of random RNA it was found
\cite{mezmul} numerically $\gamma_{\rm gl}\simeq 2/3$, and in the
recent work \cite{weise} the analytical estimate $\gamma_{\rm
gl}\simeq 5/8$ has been obtained for the same system via the
renormalization group procedure.

Consider now the topology of the bonds formed in the AB--copolymer. One can easily
see that A--B bonds separate the independent regions along the chain. Let us start
moving along the chain from the left end and find the first A--B bond. As the arcs
cannot intersect each other (see \fig{fig:2}), all the monomers under the given arc
can be connected only with each other. Moreover, the A monomers in the beginning of
the chain can form only A--A bonds (as we are speaking about the {\it first} A--B
bond), and the B monomers in the rest of the chain can form only B--B bonds with
each other, too. This separation of a chain into three independent parts allows us
to write down the following Dyson--type equation for the partition function of the
whole chain:
\be
\begin{array}{r}
\disp G_2(n,m,{\bf q}|v_1,v_2,u) = G(n,{\bf q}|v_1)\, G(m,{\bf q}|v_2) +
u\sum_{i,j}G(i,{\bf q}|v_1)\, G(j,{\bf q}|v_2) \\  \disp \times G_2^{(\rm
loop)}(n-i-1,m-j-1|v_1,v_2,u) \label{eq:1}
\end{array}
\ee
where $G_2(n,m,{\bf q}|v_1,v_2,u)$ is the desired partition function (written in the
${\bf q}$--space), $G(k,{\bf q}|v)$ is the partition function of a self--associating
chain of A or B monomers of length $k$ (we assume for simplicity that the only
difference between A and B monomers is in the weight of bond formation,
respectively, $v_1$ and $v_2$), and $G_2^{(\rm loop)}(n,m|v_1,v_2,u)$ is a partition
function of a diblock--copolymer which is forced to form a loop (i.e. its end
monomers are forced to form a bond). Finally, the first term in the sum on the
r.h.s. of eq. \ref{eq:1} is due to the fact that there can be no A--B bond at all.
Note that to solve \eq{eq:1}, apart from defining the homopolymer partition
functions $G$ one should allow for some particular connection between the partition
functions of a "simple", $G_2$, and "looped", $G_2^{(\rm loop)}$, chains.

\begin{figure}[ht]
\epsfig{file=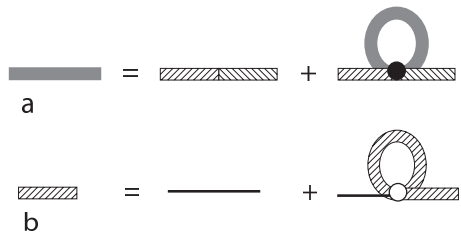,width=6cm} \caption{Diagrammatic form of Dyson--type
equations: a) for Eq.\eq{eq:1a}, b) for Eq.\eq{eq:2}. Thin lines correspond to free
propagators $g(s,{\bf q})$, thick hatched lines --- to the homopolymer propagators
$G(s, {\bf q}|v)$, and thick grey lines --- to the diblock copolymer propagators
$G_2(s_1,s_2,{\bf q}|v_1,v_2,u)$. Finally, open and filled circles correspond to the
intra-- and inter--species bonds, respectively.} \label{fig:3}
\end{figure}

Introducing the generating functions for all relevant partition functions:
\be
\begin{array}{l}
\disp G_2(s_1,s_2,{\bf q}|v_1,v_2,u) = \sum_{m=0}^{\infty}
\sum_{n=0}^{\infty}G_2(m,n,{\bf q}|v_1,v_2,u)s_1^m s_2^n; \\
\disp G(s,{\bf q}|v) = \sum_{m=0}^{\infty}G(m,{\bf q}|v)s^m; \\
\disp G_2^{(\rm loop)}(s_1,s_2|v_1,v_2,u) = \sum_{m=0}^{\infty}
\sum_{n=0}^{\infty}G_2^{(\rm loop)}(m,n|v_1,v_2,u)s_1^m s_2^n
\end{array}
\label{eq:gen}
\ee
we can rewrite equation (\ref{eq:1}) in the following form:
\be
\begin{array}{r}
\disp G_2(s_1,s_2,{\bf q}|v_1,v_2,u)=G(s_1,{\bf q}|v_1)G(s_2,{\bf q}|v_2) + u s_1
s_2 G(s_1,{\bf q}|v_1) G(s_2,{\bf q}|v_2) \medskip \\ \disp \times G_2^{(\rm
loop)}(s_1,s_2|v_1,v_2,u) \label{eq:1a}
\end{array}
\ee
(see \fig{fig:3}a for the diagrammatic form of this equation). Moreover, the
partition functions of a homopolymer forming clover--leaf structures (i.e.,
structures without pseudoknots) are known (see, for example,
\cite{erukh78,mueller,mez_pull}) to satisfy the equation
\be
G(s,{\bf q}|v)=g(s,{\bf q})+v g(s,{\bf q}) G(s,{\bf q}|v) G^{(\rm loop)}(s|v)
\label{eq:2}
\ee
(see \fig{fig:3}b for its diagrammatic form) and the function $g(s,{\bf q})$ is a
partition function of a free chain (i.e., chain without any thermoreversible bonds
or volume interactions).

The particular form of $g(s,{\bf q})$ is known to be irrelevant for long enough
chains (see, for example, \cite{Gros_Khokh}). In our case, when a chain can form
short--ranged loops, the particular form of the bare propagator influences
importantly the entropy penalties for closing a chain into a loop. Note, however,
that we introduce the loop penalties explicitly via the connection between $G$ and
$G^{(\rm loop)}$ in \eq{eq:2} and therefore we assume the form of the function
$g(s,q)$ to be irrelevant. We can thus set without the loss of generality that
\be
g(N,R)=\left(\frac{3}{2\pi
Na^2}\right)^{3/2}\exp\left(-\frac{3R^2}{N a^2}\right)
\label{eq:gauss0}
\ee
where $a$ is a mean length of a covalent bond between two subsequent monomers. After
the Fourier--Laplace transform one gets
\be
g(s,{\bf q})=\frac{1}{e^{{\bf q}^2
a^2/6}-s} \label{eq:gauss}
\ee
In what follows we chose the length units to set $a^2/6=1$. Note, that while the
Gaussian form of the free chain propagator is a matter of assumption, the value of
the propagator at $q=0$ (namely, $(1-s)^{-1}$) is independent of its particular
spatial dependence. Indeed, the partition function of a free chain with given number
of monomers $Z(N)=\int Z(N,r) d^3 r$ is by definition equal to unity (see
\cite{Gros_Khokh}). Therefore, the value of $g(q,s)$ at $q=0$, which is just
$\sum_{N=0}^\infty Z(N) s^N$, equals always $g(0,s)=1/(1-s)$.

The crucial difference between equations (\ref{eq:1a}) and (\ref{eq:2}) is that
unlikely to the latter one, the former equation is linear. This corresponds to the
aforementioned fact that the A--B bonds separate the chain into independent parts,
therefore the global structure of the bonded chain is just a linear sequence of
independent "blobs" separated by A--B bonds, only A--A and B--B bonds allowed inside
each blob. This notion leads us to understanding of the main feature of the system
under consideration: the possibility of a necklace--cactus transition. Indeed, for
$v_{1,2}\gg u$ one can neglect the alternating A--B bonds and the chain forms just
two independent "cactuses" consisting of A and B monomers, the roughness exponent
$\gamma$ being equal to 1/2 similarly to the homopolymer case. On the other hand, if
$u \gg v_{1,2}$, the alternating bonds dominate in the structure, and the chain
forms a "hairpin--like" structure with $\gamma=1$. Even a stronger conjecture is
true: if the fraction of A--B bonds is nonzero in the limit of a long chain, the
roughness exponent $\gamma$ of such a configuration is equal to 1.

We are going now to analyze the described transition in more detail. To close the
system of equations we should now address the aforementioned problem of the
connection between "simple" and "looped" partition functions. In the two following
Sections we consider, respectively, the following assumptions about this connection:
there are no loop weights (Section \ref{sect:3}); there are "ideal" loop weights
(Section \ref{sect:4}).

To be specific, in the Section \ref{sect:3} we assume the most simple possible
situation. Namely, we postulate:
\be
\begin{array}{rll}
G_2^{(\rm loop)}(s_1,s_2) & = & \disp  G_2(s_1,s_2,{\bf q}=0)=\int G_2(s_1,s_2,{\bf
r}) d{\bf r} \medskip \\ G^{(\rm loop)}(s) & = & \disp G(s,{\bf q}=0)=\int G(s,{\bf
r})d{\bf r}
\end{array}
\label{eq:nol}
\ee
Here the functions $G(s, {\bf r})$ and $G(s_1,s_2,{\bf r})$ are
the partition functions of the homopolymer and copolymer chains in
the ${\bf r}$--space. The assumption (\ref{eq:nol}) corresponds
obviously to omitting any entropic penalties for loop formation.
We suppose that such an approximation can be valid at least for
low enough temperatures, for the justification of this assertion
see the beginning of the Section \ref{sect:3}.

In the Section \ref{sect:4} we assume more natural connection between the partition
functions under discussion:
\be
\begin{array}{rll}
G_2^{(\rm loop)}(s_1,s_2) & = & \disp  \frac{1}{(2\pi)^3}
\int G_2(s_1,s_2,{\bf q})d{\bf q}=G_2(s_1,s_2,{\bf r}=0) \medskip \\
G^{(\rm loop)}(s) & = & \disp \frac{1}{(2\pi)^3} \int G(s,{\bf q})d{\bf q}=G(s,{\bf
r}=0)
\end{array}
\label{eq:idloop}
\ee
where the r.h.s. is written also in terms of the partition function in the ${\bf
r}$--space. Thus, we simply assume here that the partition function of a loop (i.e.
chain, actually bonded at the end points) equals the partition function of a similar
chain without end-to-end bond, but whose end monomers are situated at the same point
in space. This assumption on the first glance looks almost obvious and it is hard to
understand why it could fail. Note, however, that the above equations do not allow
for volume interactions, i.e. interactions of non-saturating nature between
monomers, situated far from each other along the chain, but close to each other in
the real space. These interactions, however, do not influence the topology of
possible cloverleaf conformations. Having in mind the discussion of the major
factors, governing the the problem of a RNA--type folding, which we have presented
in Section \ref{sect:1}, one can hope to take the volume interactions into account
via altering the loop weight function, or, in our formalism, by choosing the proper
connection between the "looped" and "non--looped" partition functions in the form
different from the ideal one \eq{eq:idloop}.

\section{No loop weights}
\label{sect:3}

In this Section we calculate the partition function of the chain with no penalties
for the loop formation. To justify this simplification let us remind that the loop
weight is supposed to be relevant only if the loops formed in the system are long
enough. Indeed, the loop weight, i.e. the probability that a polymer chain of length
$l$ forms a loop, is, at least in the first approximation, proportional to
$R^{-D}_l$, were $R_l$ is the typical end-to-end distance of such a chain, and $D$
is the space dimensionality. In turn, $R_l$ is proportional to some positive power
of $l$ depending on the state of the system. Therefore, the loop weights are of
order of unity for short loops and much less than unity for long ones. In other
words, the loop weights effectively suppress the long loops. If such loops are
already suppressed (which is indeed the case if $u,v_{1,2} \gg 1$, when almost all
the monomers are involved into bonds), the particular form of loop weights is not
much relevant. Hence, for large enough values of association constants (i.e., for
low enough temperature) the no-loop-weights approximation seems to be realistic.

Keeping this in mind, we can simplify equations \eq{eq:1a} and \eq{eq:2} as follows.
We substitute the total partition functions $G({\bf q}=0)$ and $G_2({\bf q}=0)$
instead of loop ones $G^{(\rm loop)}$ and $G_2^{(\rm loop)}$ and then set ${\bf q}$
equal to zero. We arrive therefore to a set of two algebraic equations:
\be
\begin{array}{rcl}
G(s|v) & = & g(0,s)+ v g(0,s)s^2 G^2(s|v) \medskip \\
G_2(s_1,s_2|v_1,v_2,u) & = & G(s_1|v_1)G(s_2|v_2) + u s_1 s_2 G(s_1|v_1)G(s_2|v_2)
G_2(s_1,s_2|v_1,v_2,u)
\end{array}
\label{eq:noloop}
\ee
The solutions to \eq{eq:noloop} read:
\be
G(s,v)=\frac{1-s-\sqrt{(1-s)^2-4 v s^2}}{2 v s^2}
\label{eq:nol1}
\ee
for the homopolymer partition function (we have chosen the branch of solutions which
approaches $G=(1-s)^{-1}$ as $v \to 0$); and
\be
\begin{array}{rcl}
G_2(s_1,s_2) & = & \disp \frac{1}{G^{-1}(s_1,v_1)G^{-1}(s_2,v_2)-u s_1 s_2} \medskip \\
& = & \left(\dfrac{4 v_1 v_2 s^2_1 s^2_2}{\left(1-s_1-\sqrt{(1-s_1)^2-4 v_1
s^2_1}\right) \left(1-s_2-\sqrt{(1-s_2)^2-4 v_2 s^2_2}\right)} -u s_1
s_2\right)^{-1}
\end{array}
\label{eq:nol2}
\ee
for the block-copolymer partition function.

Note that though we have introduced the generating functions $G(s)$ and
$G_2(s_1,s_2)$ as purely formal objects, one can ascribe a physical meaning to the
result obtained in \eq{eq:nol2}. Indeed, imagine a polydisperse system of
AB--copolymers in the equilibrium with a reservoir of A-- and B--monomers. The
length of A-- and B--blocks will then be controlled by corresponding fugacities
$s_1$ and $s_2$, and the partition function of such a block--copolymer is given by
\eq{eq:nol2}. An important note is that this situation would be fully annealed with
respect to the lengths of the blocks, while in most of (at least biologically
relevant) experimental cases the length and primary structure of a polymer are
quenched. In what follows we are going to consider the quenched case when the length
of the blocks is fixed and does not depend on the particular form of the interaction
between monomers.

It seems now instructive to study the analytic structure and determine singularities
of \eq{eq:nol2}, because the particular form of these singular points is responsible
for the equilibrium conformations of the system under consideration. Indeed, one can
easily see that there are three possible singularities of \eq{eq:nol2}: two
square--root singularities at the points
\be
v_1=\left(\frac{1-s_1}{2 s_1}\right)^2; \quad v_2=\left(\frac{1-s_2}{2 s_2}\right)^2
\label{eq:sqroot}
\ee
and the simple pole singularity at the point
\be
\frac{4 v_1 v_2 s_1 s_2}{u}=\left(1-s_1-\sqrt{(1-s_1)^2-4 v_1 s^2_1}\right)
\left(1-s_2-\sqrt{(1-s_2)^2-4 v_2 s^2_2}\right)
\label{eq:pole}
\ee

The square--root singularity corresponds (see, for example, \cite{mueller}) to the
formation of a usual randomly branched clover--leaf structure with $\gamma=1/2$ due
to the self--association of monomers of type A (or B), while the simple pole signals
the formation of a linear ($\gamma=1$) double--folded conformation due to the
alternating A--B association. Thus, if the square--root singularity "wins" (i.e.
becomes more important), the global conformation of a chain is a randomly branched,
while if the pole singularity "wins", the global conformation is linear. One can
imagine also a situation when (say, for a very long A--block and short B--block)
there can arise also a mixed situation, when both singularities are of equal
importance or, saying it in more physical words, when within a single polymer there
is a phase equilibrium between a necklace (double--stranded) phase and a homopolymer
(in this case -- A) cloverleaf phase. However, as we show below, this situation
actually never exists in the particular system under discussion.

If a system is controlled by one single activity $s$, the meaning of "winning
singularity" is well-defined: in the thermodynamic limit the singularity closest to
$s=0$ is dominant. In the case under consideration we have two independent
activities $s_1$ and $s_2$, which makes the situation somewhat more tricky. However,
one can handle this problem in the case of two activities considering the
thermodynamic limit: $m=\alpha N$; $n=(1-\alpha)N$ when $N \to \infty$ and
$\alpha={\rm const}$.

Here we present only the results, while the details of the algorithm we are using
are explained in the Appendix A, and the details of particular computations -- in
the Appendix B.

Depending on the particular value of the association constants $\{u, v_1, v_2\}$
either the square--root or the pole singularity become dominant. So, we have two
options:

If $u<u_{\rm tr}=\sqrt{v_1 v_2}$, then the square root singularity is dominant and
the system is in the cloverleaf phase with the free energy per link of the chain $f$
asymptotically (when $N \to \infty$) equal to
\be
f_{\rm clov}=\ln s^*_{\rm sqr} = -\alpha\ln(1+2\sqrt{v_1})-(1-\alpha)
\ln(1+2\sqrt{v_2}) \label{eq:freeclover}
\ee
(See Eq.(\ref{eq:free1})).

If $u>u_{\rm tr}$, then the pole singularity is dominant and the polymer is in the
necklace conformation. The corresponding free energy reads
\be
f_{\rm neck}=\ln s^*_{\rm pole} = -\alpha \ln(1+ 2
\sqrt{v_1}\cosh(\vartheta-\xi))-(1-\alpha)\ln(1+2 \sqrt{v_2} \cosh(\vartheta+\xi))
\label{eq:freeneck}
\ee
(see (\ref{eq:free2})), where
\be
\vartheta = \frac{1}{4}\ln\frac{u^2}{v_1 v_2} \label{eq:theta1}
\ee
and $\xi$ in \eq{eq:freeneck} is the solution of the equation
\be
4\sqrt{v_1 v_2} \sinh 2\xi + 4(1-2\alpha) \sqrt{v_1 v_2} \sinh 2 \vartheta -\alpha
\sqrt{v_1} \sinh(\vartheta-\xi) +(1-\alpha) \sqrt{v_2} \sinh(\vartheta+\xi)=0
\label{eq:ksi2}
\ee
on the interval $(-\vartheta, \vartheta)$.

\begin{figure}[ht]
\epsfig{file=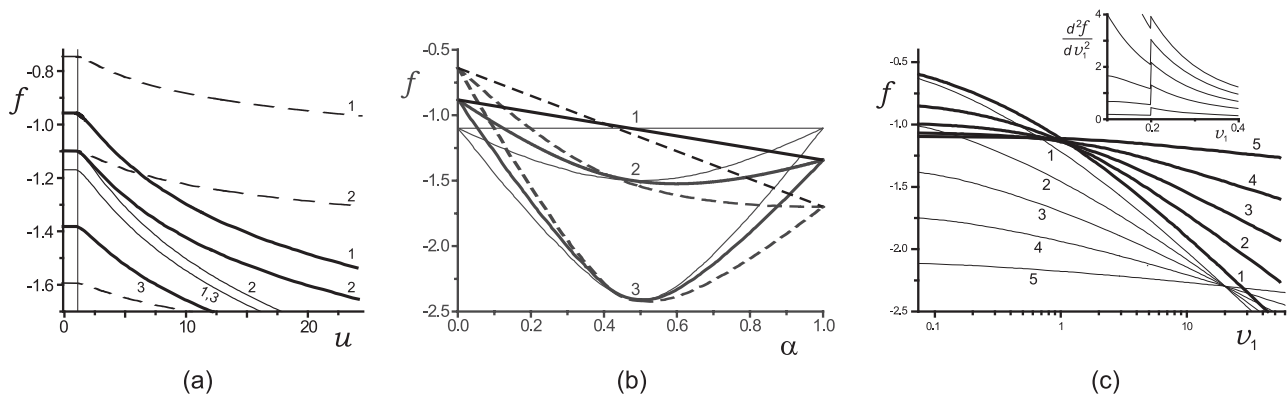,width=17cm} \caption{Free energy of the system defined by
(\ref{eq:freeclover}) and (\ref{eq:freeneck}) as a function of $u$ (a), $v_1$ (b)
and $\alpha$ (c). (a, b): Thin lines correspond to $v_1=v_2=1$, thick lines -- to
$v_2=2,\, v_1 =0.5$, dashed lines -- to $v_2=5,\, v_1=0.2$. (a) Curves corresponding
to $\alpha =$ 0.1, 0.5, and 0.9 are marked by letters {\it a}, {\it b}, and {\it c},
respectively. (b) Curves corresponding to $u = $ 1 (the transition point), 10, and
100 are marked by  letters {\it a}, {\it b}, and {\it c}, respectively. (c) Thick
lines correspond to $u=2,\, v_2=1$, thin lines, both in the main figure and the
inset -- to $u=2,\, v_2=20$. Curves marked by 1, 2, 3, 4, and 5 correspond to
$\alpha =$ 0.9, 0.7, 0.5, 0.3, and 0.1, respectively. Note that the thin lines
intersect at the same point $v_1=v_2$ which belongs to the cloverleaf phase. This is
not true for the thick lines since the point $v_1=v_2$ belongs to the necklace phase
in this case.} \label{fig:4}
\end{figure}

In Figure \ref{fig:4} we have plotted the free energy of the system defined by
equations (\ref{eq:freeclover}) and (\ref{eq:freeneck}) as a function of variables
$u$, $v_1$ and $\alpha$ --- see Figs. \ref{fig:4}a,b,c, respectively.

The free energy does not depend on $u$ in the cloverleaf phase and it diverges
quadratically at the transition point from the cloverleaf value, thus giving rise to
the discontinuity of its second derivative. For $f(v_1)$ dependencies this
discontinuity is less evident. To make it more clear we have plotted in the inset
the function $\frac{d^2f(v_1)}{dv_1^2}$. The transition from the cloverleaf to the
necklace phase is, thus, a 2nd order phase transition. Moreover, it is very
important that the function $f(\alpha)$ is always a convex curve. This fact provides
a support for our assertion that there is no any phase separation inside the system
under consideration. Indeed, if there were the regions where the free energy as a
function of $\alpha$ is concave, the system would have minimized its energy in a
usual way, by separating into two necklace-phases with different effective values of
$\alpha$. However, as we see, it is not the case. Together with our consideration in
the end of Appendix B where we prove that there is no necklace--cloverleaf phase
coexistence in the system, this leads us to the conclusion that the homogenous phase
in the system under consideration for any $\alpha$ is always more stable (i.e., has
lower free energy) then two phases.

\section{Ideal loop weights}
\label{sect:4}

In this section we analyze the exact solution of \eq{eq:1a} and \eq{eq:2} with the
loop weights defined by \eq{eq:idloop} and with the propagator of a free chain given
by \eq{eq:gauss}. Substituting \eq{eq:gauss} and \eq{eq:idloop} into \eq{eq:2} one
gets a nonlinear integral equation which is (see \cite{erukh78,mez_pull}) solvable
by means of integration over ${\bf q}$. One gets
\be
G(s,v,{\bf q})=\frac{1}{e^{{\bf q}^2}-X(s,v)} \label{eq:single}
\ee
(compare to the "bare" function \eq{eq:gauss} and recall that we have set $a^2/6=1$)
where $X(s,v)$ satisfies the following equation
\be
(4\pi)^{3/2}X(X-s)=v s^2 \text{Li}_{3/2}(X)
\label{eq:polylog}
\ee
and $\text{Li}_{3/2}(X)$ is a polylogarithm function of order 3/2. Equation
\eq{eq:polylog} always has a solution for small enough $s$. When $s$ is increasing,
\eq{eq:polylog} loses the solution at the point defined by the derivative of
\eq{eq:polylog} with respect to $X$:
\be
(4\pi)^{3/2}X(2X-s)=v s^2 \text{Li}_{1/2}(X)
\label{eq:dpolylog}
\ee
Let us note that at this point the solution $X(s,v)$ (and, therefore $G(s,v,{\bf
q})$) has a square root singularity similar to that of equation \eq{eq:sqroot}.
Substituting \eq{eq:single} together with the first equation of \eq{eq:idloop} into
\eq{eq:1a}, one arrives at the partition function of a diblock copolymer with ideal
loop weights:
\be
A_2(s_1,s_2)=\dfrac{1}{(2\pi)^3}\int G_2 (s_1,s_2,{\bf q})d{\bf q} = \left(
\dfrac{X(s_1,v_1)-X(s_2,v_2)}{\dfrac{X(s_1,v_1)-s_1}{v_1 s_1^2}
-\dfrac{X(s_2,v_2)-s_2}{v_2 s_2^2}} -u s_1 s_2 \right)^{-1} \label{eq:block}
\ee
The equation \eq{eq:block} has a simple pole singularity at the point when the whole
expression in brackets in the r.h.s. equals zero. Note that there is no singularity
due to the denominator of the ratio inside the brackets, since the nominator and
denominator always tend to zero simultaneously, while the ratio as a whole stays
finite. Thus, one sees again that the conformation of the block--copolymer under
discussion is governed by an interplay between a square root singularity
(corresponding to the cloverleaf phase) and a simple pole (corresponding to the
necklace phase).

Similarly to the previous section, we sketch here the results for the free energy
and the transitions in the system, putting all the details of computations into
Appendix C.

\begin{figure}[ht]
\epsfig{file=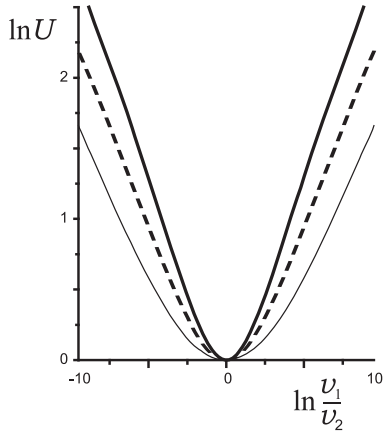,width=5cm} \caption{Phase diagram of the
cloverleaf--necklace phase transition with ideal loop weights. Thin, dashed and
thick lines correspond to the product $v_1 v_2 = $ 100, 10, and 1, respectively.}
\label{fig:5}
\end{figure}

The transition point now is given by
\be
U = \frac{u}{\sqrt{v_1 v_2}} = \frac{\sqrt{v_1 v_2} (X^*_1 -
X^*_2)}{v_2 s^*_2 \left(X^*_1 (s^*_1)^{-1} - 1\right) - v_1 s^*_1
\left(X^*_2 (s^*_2)^{-1} - 1\right)}
\label{eq:tranideal}
\ee
where $(s^*_1, X^*_1)$ and $(s^*_2, X^*_2)$ are the common solutions of equations
(\ref{eq:polylog}) and (\ref{eq:dpolylog}) for $v=v_1$ and  $v_2$, respectively. In
\fig{fig:5} we plot the phase transition line $U_{\rm tr}(v_1,v_2)$ as a function of
$v_1/v_2$ for different values of the mean homopolymer association constant
$\sqrt{v_1 v_2}$. Note that in this case the dependence $U_{\rm tr}(v_1/v_2)$ is
more rich than in the no-loop-weight case, where the transition point was at $U_{\rm
tr}=\frac{u_{\rm tr}}{\sqrt{v_1 v_2}}=1$ independently of the ratio of the
homopolymer association constants. In particular, this nontrivial dependence means
that there is a possibility of reentrant cloverleaf--necklace--cloverleaf transition
in the system under discussion. Indeed, it is natural to assume (compare to
\cite{erer,mtglob,mterJCP}) that the association constants depend on temperature in
a usual Boltzmann way:
\be
u = u_0 \exp(-\epsilon_0/(k_B T)); \quad v_{1,2}=v_0^{1,\,2}
\exp(-\epsilon_{1,2}/(k_B T)) \label{eq:boltzmann}
\ee
This means that the change of temperature in any particular system with given
chemical structure and given values of $u_0,\,v_0^{1,\,2},\,\epsilon_{0,1,2}$
corresponds to a movement along some straight line on the log--log plot in
\fig{fig:5}. We easily see that such a straight line can easily have at least two
intersections with the phase transition line thus giving rise to the aforementioned
reentrant transition \footnote{Note that, generally speaking, the mean homopolymer
constant $\sqrt{v_1 v_2}$ is changing itself when the temperature changes. However,
this change only slightly perturbs the transition curve and does not influence the
aforementioned result.}.

Moreover, the free energy of the system with ideal loop weights reads
\be
f = \alpha \ln s_1^{\rm a} + (1-\alpha) \ln s_2^{\rm a} \label{eq:freeideal}
\ee
where in the cloverleaf phase the acting singularity values $s_{1,2}^{\rm a}$ are
equal to the $s_{1,2}^*$ defined in \eq{eq:tranideal}, while in the necklace phase
they are the solutions of the set of equations (\ref{eq:polylog}),
(\ref{eq:idpole}), and (\ref{eq:idpolecondition}) lying within the rectangle $0 <
s_1^{\rm a} \leq s_1^*; 0 < s_2^{\rm a} \leq s_2^*$.

\begin{figure}[ht]
\epsfig{file=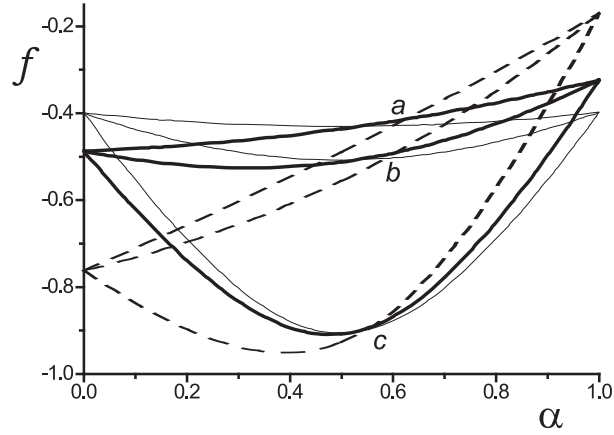,width=8cm} \caption{Free energy of the system with ideal
loop weights as a function of $\alpha$. Thin, thick, and dashed lines correspond to
$v_1=v_2=14.1$; $v_1=10,\, v_2=20$; and $v_1=4,\, v_2=50$, respectively. The curves
corresponding to $u=40$, $u=80$, and $u=400$ are marked by {\it a}, {\it b}, and
{\it c}, respectively. Note that $u=40$ (curves {\it a}) correspond to a point in
the necklace phase just above the transition. In the cloverleaf phase (see
\eq{eq:fidclover}) the free energy is linear in $\alpha$.} \label{fig:6}
\end{figure}

In \fig{fig:6} we plot this free energy $f$ as a function of
$\alpha$ for several different values of $v_1$, $v_2$ and $u$. One
can notice the similarity between this plot and \fig{fig:4}b. The
curves are once again always convex, so there is once again no
internal phase separation in the system. Moreover, one can easily
check that the necklace--cloverleaf transition is again of 2nd
order. Thus we see that the difference between no-loop-weight and
ideal-loop-weight cases is mainly qualitative, except the
peculiarities in the behavior of the transition point $U_{\rm
tr}(v_1,v_2)$ and the possibility of the reentrant transition in
the latter case.

\section{Conclusions}
\label{sect:5}

We have shown that a diblock copolymer capable of forming RNA--like structures due
to reversible bond formation can exhibit a phase transition between two different
thermodynamically equilibrium states, a "necklace" with the roughness exponent
$\gamma_{\rm nc}=1$, and a "cloverleaf" with $\gamma_{\rm rb}=1/2$. The transition
is governed by the values of equilibrium association constants, and the weights of
loop formation in the system.

We have demonstrated that a cloverleaf--necklace transition in the thermodynamic
limit is a 2nd order phase transition both in the case of no-- and ideal loop
weights (in these cases we have calculated the free energy of the system exactly).
One can expect that allowing larger penalties for loops formation, the new phase,
that of a coil, may arise in the system, similarly to what predicted in
\cite{mueller}, and the order of the necklace--cloverleaf transition itself may
change.

The interesting feature of the systems under discussion in the case of ideal loop
weights is the possibility of a reentrant necklace--cloverleaf transition for some
particular values of parameters determining the temperature dependence of the
association constants.

Moreover, the proposed approach allows to calculate the mean free energy of a frozen
ensemble of the diblock RNA--like copolymers with some given distribution of bond
lengths.

Finally, we would like to emphasize that the method of calculating
the partition function of a diblock--copolymer based on
\eq{eq:1a}, \eq{eq:2} (or equations (\ref{eq:noloop}) in the case
of no loop weights) can be straightforwardly generalized to the
case of alternating polyblock RNA--like copolymers of type ${\rm
A}_{t_1}{\rm B}_{t_2}{\rm A}_{t_3}{\rm B}_{t_4}...{\rm
A}_{t_{k-1}}{\rm B}_{t_k}$ with all lengths of blocks being
different. This situation models a particular example of an
alternating heteropolymer RNA--like molecule with quenched primary
structure defined by the sequence of length $\{t_1,t_2,...,t_k\}$.
For example, the case of a 4--block copolymers seems to be of
particular interest, as in this case one can expect the
equilibrium cross--like conformations reminiscent of that of the
tRNA.

\begin{figure}[ht]
\epsfig{file=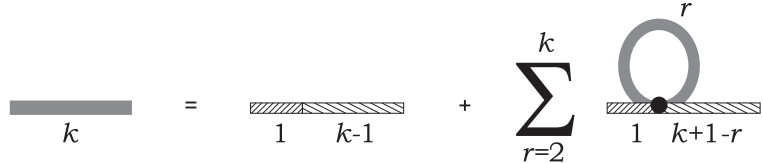,width=10cm} \caption{Diagrammatic structure of equation
\eq{eq:dyson_poly}.} \label{fig:8}
\end{figure}

The corresponding general Dyson--type equation for the generation function
$G_n(s_1,s_2,...,s_n)$, where
$$
G_n(s_1,s_2,...,s_k) = \sum_{t_1=0}^{\infty} \sum_{t_2=0}^{\infty} ...
\sum_{t_k=0}^{\infty} G_k(t_1,t_2,...,t_k) s_1^{t_1}s_2^{t_2}...s_k^{t_k}
$$
takes the following form (compare to \eq{eq:1a}):
\be
G_k(s_1,...,s_k) = G_1(s_1)\,G_{k-1}(s_2,...,s_k)+\sum_{r=2}^{k} u_{1,r}\, s_1\,
s_r\, G_1(s_1)\, G_r(s_1,...,s_r)\, G_{k+1-r}(s_r,...,s_k) \label{eq:dyson_poly}
\ee
which has the diagrammatic structure shown in \fig{fig:8} (compare to \fig{fig:3}a).
The equation \eq{eq:2}, as well as it diagrammatic structure shown in \fig{fig:3}b
remain without changes.

We believe that equation \eq{eq:dyson_poly} can be served as a basis of a
renormalization group procedure, thus connecting the exact partition function,
$G(4)$, of an 4--block copolymer with an approximative expression for $G(4)$
constructed of two diblock copolymers, $G(2)$, i.e. $G(4) \approx G(2)\times G(2)$.
It is obvious that if all coupling constants are identical, then we have a RNA--like
homopolymer and the the last relation is exact. However if A--A, B--B and A--B
constants are slightly different, we should have $G(4)-G(2)\times
G(2)=\varepsilon(u,v_1,v_2)\neq 0$. Considering $\varepsilon$ as a perturbation we
may look how $\varepsilon$ is transformed under the standard coarse--graining
procedure. This work is in progress and the corresponding results will be reported
elsewhere \cite{ta_ne}.

\begin{acknowledgments}

The authors are grateful to I.E. Erukhimovich for valuable discussions. M.V.T.
gratefully acknowledges warm hospitality he felt throughout his stay at the LPTMS
where this work was done. The work is partially supported by the grant
ACI-NIM-2004-243 "Nouvelles Interfaces des Math\'ematiques" (France). We are
grateful to S. Lando and P. Krapivsky for comments of mathematical character and for
providing us relevant references on multivariate complex analysis.

\end{acknowledgments}

\begin{appendix}

\section{Singularities of the generating function $G(s_1,s_2)$}

In this Appendix we address the procedure of calculating the asymptotic behavior of
the coefficients of a generating function (compare to \eq{eq:gen})
\be
G(s_1,s_2)=\sum_{n,m=0}^{\infty} a_{m,n} s_1^m s_2^n \label{eq:gendef}
\ee
in the limit $m+n=N\to \infty$ and $m/(m+n)\equiv\alpha={\rm const}$. The procedure
suggested below is similar to that developed in \cite{twenty} and we advise the
reader to address this reference for the accurate mathematical proofs. However, the
algorithm under discussion is not of a common use, so it seems instructive to give a
short explanation on the physical level of rigor. Note, that we pay attention only
to the exponentially increasing multipliers in the asymptotic behavior of the free
energy, thus neglecting the logarithmic corrections.

First of all, let $G(s_1,s_2)$ defined by (\ref{eq:gendef}) be an analytic function
of $s_1, s_2$ in the circle $\sqrt{|s_1|^2+|s_2|^2}<\epsilon$ for some $\epsilon>0$
near the origin $s_1=s_2=0$. Assume also, as it is usual in the statistical
mechanics, the dominant singularities of $G$ with respect to each of the variables
(for fixed another one) to be real and positive.

Now, it is known that to calculate a function $G_1^*(s)$ defined
as follows:
\be
G_1^*(s)=\sum_{i=0}^{\infty} a_{i,i}\, s^{2i} \label{eq:trace}
\ee
whose dominant singularity defines the asymptotic of $a_{i,i}$ at
$i \rightarrow \infty$, one can use the following trick (see, for
example, \cite{Lando}). Consider $G\left(sz, \frac{s}{z} \right)$
as a function of a variable $z$. If $s$ is small enough this
function is analytic with respect to $z$ inside some ring where it
can be represented by the Laurent series
\be
G\left(s z, \frac{s}{z}\right)=\sum_{k=-\infty}^{\infty} b_k(s)\, z^k
\label{eq:Laurent}
\ee
Note now that the coefficient $b_0(s)$ in this expansion is
exactly equal to the desired function $G_1^*(s)$ and therefore
this function can be evaluated using the Cauchy theorem:
\be
G_1^*(s)=\frac{1}{2 \pi i} \oint_C G\left(s z, \frac{s}{z}\right)
\, \frac{dz}{z}
\label{eq:Cauchy}
\ee
where the curve $C$ encloses the point $z=0$ exactly once and belongs to the region
of analyticity of the function $G\left(sz, \frac{s}{z}\right)$.

This consideration can be easily generalized in the following way. Suppose, we are
interested in calculating the function
\be
G^*\left(s,\frac{p}{q}\right)=\sum_{j=0}^{\infty} a_{jp,\, jq}\,
s^{j(p+q)}
\label{eq:trace2}
\ee
where $p$ and $q$ are two mutually simple natural numbers. To perform the summation
in \eq{eq:trace2}, we consider a function $G(s z^q, s z^{-p})$ and take the
coefficient of order $0$ in its Laurent series:
\be
G^*\left(s,\frac{p}{q}\right)=\frac{1}{2 \pi i} \oint_{C} G\left(s
z^q, \frac{s}{z^p}\right)\, \frac{dz}{z} \label{eq:Cauchy2}
\ee
where the integration curve belongs again to the region of analyticity of $G(s z^q,
s z^{-p})$.

This formalism allows one in principle to calculate the generating function
$G^*(s,\frac{p}{q})$ for any given ratio of block lengths of the copolymer under
discussion. However, on the one hand, the explicit calculation of $G^*$ is still
rather cumbersome, and on the other hand, it is actually not needed. What we are
really interested in is the {\it asymptotic} behavior of the coefficients of $G^*$,
or, in other words, the position and the nature of its dominant singularity (or
singularities). The task of finding these singularities can be accomplished without
explicit knowledge of $G^*$.

To do that, let us first introduce some additional definitions. Let $H$ be the 3D
surface of singular points of $G$ (it may consist of several branches, of course) in
the 4--dimensional space $({\rm Re}\, s_1, {\rm Re}\, s_2, {\rm Im}\, s_1, {\rm
Im}\, s_2)$ and $h$ be the intersection of $H$ with the "real positive plane" (i.e.,
the quarter--plane of $(s_1, s_2)$ where ${\rm Arg} (s_1)=0$ and ${\rm Arg}
(s_2)=0$). Now, the surface $H$ separates the  space $(s_1, s_2)$ into several
domains. We pay a particular attention to the shape of the domain containing the
origin ($s_1=s_2=0$). Let us call its edge $H_1$ ($H_1 \subset H$). It follows
immediately from the condition of single--variable dominant singularities to be real
and positive, that for any given absolute value $|s_1|$, $|s_2|$, the point of the
edge of the domain $H_1$ closest to the origin is the point with ${\rm Arg} (s_1) =
{\rm Arg} (s_2) = 0$, and, thus, it belongs to the curve $h$. In what follows we
can, restrict ourselves to consideration of real and positive singularities of $G$.

\begin{figure}[ht]
\epsfig{file=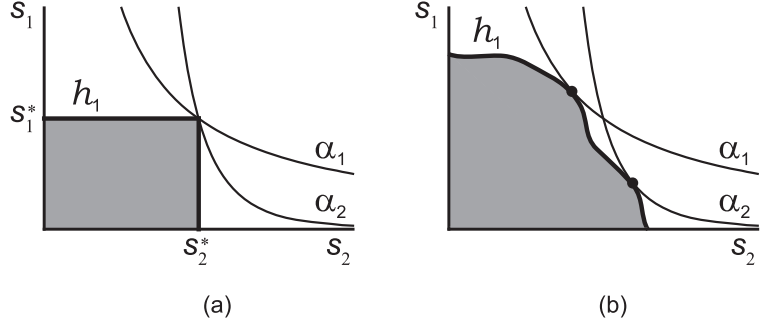,width=10cm} \caption{Examples of possible
domains of analyticity. The curves \eq{eq:s_conn} corresponding to different values
of $\alpha$ ($\alpha_1<\alpha_2$) are also shown. The touching points may (b) or may
not (a) depend on $\alpha$.} \label{fig:7}
\end{figure}

In a $(s_1,s_2)$ quarter--plane there exists some domain of analyticity of $G$ which
includes the origin, its edge is designated by $h_1$. An example of such a domain of
analyticity of $G$ in a quarter--plane $(s_1,s_2)$ including the origin, and the
edge, $h_1$ ($h_1\subset H_1$), of this domain are shown in \fig{fig:7}.

Let us fix now some "direction" $\alpha = \frac{p}{p+q}$. We are interested in
calculating the generating function $G^*(s,\alpha)$ corresponding to this particular
direction as defined by (\ref{eq:trace2}). We are thus to integrate the function
$G(s z^q, s z^{-p})$ as prescribed in (\ref{eq:Cauchy2}). By fixing some particular
value of $s$, we define therefore a connection between the values of $s_1 = s z^q$
and $s_2 = s z^{-p}$ in the integrand as follows:
\be
s_1^p s_2^q = (s^p z^{pq}) (s^q z^{-pq})=s^{p+q} \quad ({\rm or}\;
s_1^{\alpha} s_2 ^{1-\alpha} = s) \label{eq:s_conn}
\ee
Obviously, the corresponding line in the $(s_1,s_2)$ plane always intersects our
domain of analyticity if $s$ is small enough. Therefore, for small $s$ there is no
problem to choose the appropriate circle of integration $|z|={\rm const}$. One
should just choose $|z|$ in such a way that the point $(s |z|^q, s |z|^{-p})$
belongs to the domain of analyticity. However, when one increases $s$, the
intersection between this domain and the line (\ref{eq:s_conn}) may disappear at
some point. In terms of the $z$ plain, where the integration prescribed in
(\ref{eq:Cauchy2}) is to be accomplished, it means that two singularities of the
integrand, one being inside the integration curve, and another -- outside, approach
each other, and finally merge. Note, that they do merge indeed rather then just have
equal absolute values since we have assumed these singularities to be real and
positive. Thus, we see that the integral in (\ref{eq:Cauchy2}) has a singularity
exactly at this point.

Thus, summarizing the said above, we suggest the following procedure to find the
singularities of the function $G^*(s,\alpha)$:
\begin{enumerate}
\item Find the lines of singular points $h$ of the initial function $G(s_1,s_2)$ in
the real plane;
\item Define the domain of analyticity $D$ which includes the origin $s_1=s_2=0$ and
find its edge $h_1$;
\item For each value of $0<\alpha<1$ find the value of $s$ for which the
intersection of $D$ and the curve (\ref{eq:s_conn}) ceases to exist for the first
time. If $h_1$ is a smooth curve it means just finding a common tangent line of
$h_1$ and (\ref{eq:s_conn}), if there is a finite number of peculiar points on it
(say, kinks or intersections of the branches), it includes also finding the
conditions that (\ref{eq:s_conn}) contains these particular points.
\end{enumerate}

\section{Calculation of the free energy in the no-loop-weights limit.}

In this Section we apply the strategy presented in the Appendix A for calculating
the free energy corresponding to the grand partition function (\ref{eq:nol2}), i.e.
that of the diblock copolymer with no loop weights.

\begin{figure}[ht]
\epsfig{file=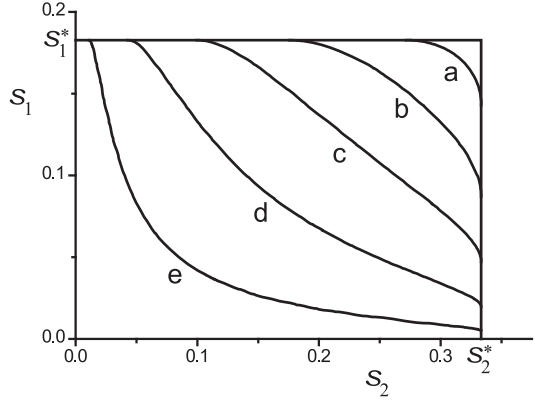,width=7cm} \caption{Examples of the analyticity domains of
(\ref{eq:nol2}) for $v_1=1, v_2=5$. Rectangular domain with borders $s_{1,2}^*=(1+2
\sqrt{v_{1,2}})^{-1}$ corresponds to values of $u$ below the necklace
cloverleaf-transition (i.e., $u<u_{\rm tr}=\sqrt{5}$). The curves marked by {\it a},
{\it b}, {\it c}, {\it d}, {\it e} are the pole lines corresponding to $u =$ 5, 10,
20, 50, and 200, respectively.} \label{fig:9}
\end{figure}

As prescribed by the suggested procedure, we should first find the domain of
analyticity which includes the point $s_1=s_2=0$. In \fig{fig:9} we have drawn these
regions in the $(s_1,s_2)$ plane for some particular sets of $v_1,v_2$ and different
values of $u$. The edges of the region correspond to singularity lines defined by
(\ref{eq:sqroot}) and (\ref{eq:pole}). The square root singularities correspond to
some straight lines $s_1=|s_1^{\rm sqr}|$ and $s_2=|s_2^{\rm sqr}|$ (where one
should choose the solution of (\ref{eq:sqroot}) with the lowest absolute value), so
for any $(s_1,s_2)$ with absolute values the rectangle $(0 \leq |s_1|<|s_1^{\rm
sqr}|,0 \leq |s_2|<|s_2^{\rm sqr}|)$ the square roots in (\ref{eq:nol2}) are both
analytical functions of their respective variables. Moreover, the pole singularity
(\ref{eq:pole}) defines some other curve in the $(s_1,s_2)$ plane. This curve can
either have no branch within the rectangle described above, and than this rectangle
is the desired domain of analyticity itself, or it can intersect the rectangle (as
shown by lines {\it a}-{\it e} in \fig{fig:9}, and then one should take the part
below and to the left of it as a desired domain of analyticity.

The further calculations can be much simplified by introducing the new variables
$$
t_{1,2}=2\sqrt{v_{1,2}}\,\frac{s_{1,2}}{1-s_{1,2}}
$$
The square root singularities now correspond just to $t_{1,2}=1$. The equation for
the pole singularity (\ref{eq:pole}) can be rewritten as follows:
\be
U=\frac{u}{\sqrt{v_1 v_2}}= \frac{t_1
t_2}{(1-\sqrt{1-t_1^2})(1-\sqrt{1-t_2^2})}
\label{eq:polet}
\ee
Now, since the function
$$
f(t)=\frac{t}{1-\sqrt{1-t^2}}
$$
decreases monotonously from $+\infty$ to 1 with $t$ increasing from 0 to 1, it is
obvious that the solution of (\ref{eq:polet}) in the square $0 \leq t_1 \leq 1, \; 0
\leq t_2 \leq 1$ exists for and only for $U \geq 1$, i.e. for $u \geq u_{\rm tr} =
\sqrt{v_1 v_2}$. Moreover, we easily see that for $u< u_{\rm tr}$ when the region of
analyticity is just the rectangle shown in \fig{fig:6}, the corridor of analyticity
for the integrand in the integral expression for $G^*$ (\ref{eq:Cauchy2}) disappears
exactly at the point when the line (\ref{eq:s_conn}) passes through the point
$(s_1^{\rm sqr},s_2^{\rm sqr})$:
\be
s^*_{\rm sqr}=(s_1^{\rm sqr})^{1-\alpha}(s_2^{\rm sqr})^{\alpha}
\label{eq:NOLclover}
\ee
The free energy per link of the chain is given therefore by
\be
f_{\rm clov}=\ln s^*_{\rm sqr}= -\alpha \ln(1+2\sqrt{v_1})-
(1-\alpha) \ln(1+2 \sqrt{v_2}) \label{eq:free1}
\ee
and the singularity of $G^*$ is in this case of square root nature which justifies
our assertion that for $u<u_{\rm tr}$ the polymer is in the cloverleaf state.

In the case when $u>u_{\rm tr}$ the region of analyticity of (\ref{eq:Cauchy2})
disappears when the curves (\ref{eq:s_conn}) and (\ref{eq:polet}) have a common
tangent line. To proceed further let us introduce the new variables $\phi_{1,2}$ as
follows
\be
\phi_i = \arcsin t_i; \quad s_i=\frac{\sin \phi_i}{\sin\phi_1 + 2
\sqrt{v_i}}; \quad i=1,2. \label{eq:phi}
\ee
The equations (\ref{eq:s_conn}) and (\ref{eq:polet}) can be now
rewritten as follows
\be
U^{-1} = \tan \frac{\phi_1}{2} \tan \frac{\phi_2}{2};
\label{eq:phiU}
\ee
and
\be
s = \left(1+2\sqrt{v_1}/\sin \phi_1\right)^{-\alpha}
\left(1+2\sqrt{v_2}/\sin \phi_2\right)^{\alpha-1} \label{eq:phis}
\ee
were $\phi_{1,2} \in [0,\pi/2]$.

The condition to have the common tangent line gives rise to the additional equation
\be
\frac{\sin \phi_1}{\sin{\phi_2}}=\frac{1-\alpha}{\alpha}\,
\sqrt{\frac{v_2}{v_2}} \frac{\sin\phi_1 \cos \phi_2 (\sin \phi_1
+2\sqrt{v_1})}{\sin \phi_2 \cos \phi_1 (\sin \phi_2 +2\sqrt{v_2})}
\label{eq:tangent}
\ee

One can easily exclude two of the three unknown variables ($s,\phi_1,\phi_2$) from
equations (\ref{eq:phis})--(\ref{eq:tangent}), and arrive finally to the following
equation
\be
4\sqrt{v_1 v_2} \sinh (2\xi) + 4(1-2\alpha) \sqrt{v_1 v_2} \sinh
(2 \vartheta) -\alpha \sqrt{v_1} \sinh(\vartheta-\xi) +(1-\alpha)
\sqrt{v_2} \sinh(\vartheta+\xi)=0
\label{eq:ksi}
\ee
for a new variable $\xi=-\frac{1}{2}\ln U - \ln \tan\frac{\phi_1}{2}$ with
$\vartheta=-\frac{1}{2}\ln U$ and additional condition $-\vartheta \leq \xi \leq
\vartheta$. It is easy to check that for any $\vartheta > 0$ (i.e., for any
$u>u_{\rm tr}$) the left hand side of this equation is an increasing function of
$\xi$ and changes the sign inside the interval $(-\vartheta,\vartheta)$. The
equation (\ref{eq:ksi}) has therefore exactly one solution with respect to $\xi$.
The fact that this solution always stays within the interval
$(-\vartheta,\vartheta)$ and never riches its limits means that the dominant
singularity in the necklace phase is always (independently of $u$ and $\alpha$) a
regular point of the pole line, not the intersection of the pole and square--root
line. This means that there is no coexistence of the necklace and cloverleaf phases
in the system for any $u$ and $\alpha$. Moreover, after substituting this solution
$\xi^*$ into the original equation for the singularity of $s$ (\ref{eq:phis}) we
obtain the final expression for the free energy in this (necklace) phase:
\be
f_{\rm neck}=\ln s^*_{\rm pole}=-\alpha \ln(1+ 2
\sqrt{v_1}\cosh(\vartheta-\xi^*))-(1-\alpha)\ln(1+2 \sqrt{v_2}
\cosh(\vartheta+\xi^*)) \label{eq:free2}
\ee

\section{Calculation of the free energy in the ideal--loop--weights limit}

In this section we apply the procedure suggested in the Appendix A to the
calculation of the free energy of the block--copolymer with "ideal" loop weights. To
that end we investigate the singularities of the grand partition function
(\ref{eq:block}) with $X_1, X_2$ defined by equation (\ref{eq:polylog}).

As we have mentioned already, there are two possible types of singularities of this
partition function. The first type corresponds to the singularities of the
homopolymer partition function, is of square-root nature and corresponds to solution
of equations (\ref{eq:polylog}) and (\ref{eq:dpolylog}). The second one is a pole
singularity corresponding to the zero of the denominator in the integrand on the
r.h.s. of (\ref{eq:block}):
\be
(X_1-X_2)\left(\dfrac{X_1-s_1}{v_1 s_1^2} -\dfrac{X_2-s_2}{v_2
s_2^2}\right)^{-1} - u s_1 s_2=0
\label{eq:idpole}
\ee
where $s_i$ and $X_i$ are connected via the equation (\ref{eq:polylog}). Contrary to
the first one, this singularity is of collective nature, i.e. it depends
simultaneously on the values of both $s_1$ and $s_2$, and it also depends on the
value of $u$.

Moreover, in full analogy with the no-loop-weights case, the square--root
singularities define for small values of $u$ the region of analyticity of a
rectangular shape in the $(s_1,s_2)$ plane. In this case (i.e. below the
necklace--cloverleaf transition) the acting singularity point is, independently of
$\alpha$, just the intersection point of the two square-root singularity lines, and
the free energy equals to
\be
f_{\text{\rm clov}}(v_{1,2},\alpha)=- \alpha \ln s_1^*  -
(1-\alpha) \ln s_2^*
\label{eq:fidclover}
\ee
(compare to \eq{eq:free1}) where $s_{1,2}^*$ are the solutions of
(\ref{eq:polylog}) and (\ref{eq:dpolylog}) with $v=v_1, v_2$.

Now, with increasing $u$ the line of poles enters the rectangle of analyticity. It
first appears at the intersection point of two square--root singularity lines, thus
giving rise to the following equation for the necklace--cloverleaf transition point:
\be
(X_1^*-X_2^*)\left(\dfrac{X_1^*-s_1^*}{v_1 (s^*_1)^2}
-\dfrac{X_2^*-s_2^*}{v_2 (s^*_2)^2}\right)^{-1} = u_{\text{tr}}
s^*_1 s^*_2
\label{eq:idutran}
\ee
where we have inserted the solutions of the set (\ref{eq:polylog}),
(\ref{eq:dpolylog}) into (\ref{eq:idpole}). This equation leads after some algebra
to \eq{eq:tranideal}.

To investigate properly the phase above the transition point (i.e., for $u>u_{\rm
tr}$) we have: i) to find the common tangent lines of the pole line
(\ref{eq:idpole}) and the line (\ref{eq:s_conn}), and ii) to check whether there is
a possibility for the acting singularity to lie at the intersection of the pole line
and the square--root singularity line, i.e. to check whether there is a kink at this
intersection. To do that we should first of all find the derivative of the pole
line. The direct calculation gives:
\be
ds_1\,\left[X'_1\left(1-\dfrac{u s_2}{v_1 s_1} \right)+u s_2
\left(\dfrac{X_2-s_2}{v_2 s_2^2}+ \dfrac{X_1}{v_1 s_1^2}\right)\right]=
ds_2\,\left[X'_2\left(1-\dfrac{u s_1} {v_2 s_2} \right)+u s_1 \left(
\dfrac{X_1-s_1}{v_1 s_1^2} + \dfrac{X_2}{v_2 s_2^2} \right) \right]
\label{eq:idpolederiv}
\ee
where $X'_{1,2}$ are the derivatives of $X_{1,2}$ with respect to
$s_{1,2}$:
\be
X'_i  = \frac{dX_i}{ds_i}=X_i\frac{(4\pi)^{3/2} X_i + 2v_i s_i
\text{Li}_{3/2}(X_i)}{(4\pi)^{3/2}X_i(2X_i-s_i)-v_i s_i^2
\text{Li}_{1/2}(X_i)};\;\; i=1,2. \label{eq:xderiv}
\ee
and the auxiliary conditions (\ref{eq:polylog}) are taken into
account.

Now, let us notice that along the square--root singularity line (say, for
$s_1=s_1^*$), the corresponding derivative ($X'_1$ in this case) diverges, while, if
we are not approaching the point $(s_1^*,s_2^*)$, all the other coefficients in
(\ref{eq:idpolederiv}) remain finite. We conclude immediately, that to comply this
equation, the derivative $\frac{ds_2}{ds_1}$ along the pole line should also diverge
in the vicinity of the point $s_1=s_1^*$. Thus, the pole line has a vertical
asymptote at $s_1=s_1^*$ and there is no kink at the intersection of the pole and
square root lines (the case $s_2=s_2^*$ is absolutely similar). We see therefore
that the acting singularity never lies at this intersection. This proves, similarly
to the no--loop--weight case, that there is no possibility of the intramolecular
necklace--cloverleaf phase separation in the system.

To calculate the free energy of the necklace phase as a function of the composition
variable $\alpha$ we should now equalize the pole line derivative
(\ref{eq:idpolederiv}) and the derivative of the line (\ref{eq:s_conn}) which is
given by
\be
\frac{ds_1}{ds_2}=-\frac{1-\alpha}{\alpha}\,\frac{s_1}{s_2}
\label{eq:sconderiv}
\ee

Collecting equations (\ref{eq:idpolederiv}) and
(\ref{eq:sconderiv}) together, we get
\be
\alpha s_2 \left[X'_2\left(1-\dfrac{u s_1} {v_2 s_2} \right)+u s_1 \left(
\dfrac{X_1-s_1}{v_1 s_1^2} + \dfrac{X_2}{v_2 s_2^2} \right) \right]= (\alpha-1)s_1
\left[X'_1\left(1-\dfrac{u s_2}{v_1 s_1} \right)+u s_2 \left(\dfrac{X_2-s_2}{v_2
s_2^2}+ \dfrac{X_1}{v_1 s_1^2}\right)\right] \label{eq:idpolecondition}
\ee
This equation, together with (\ref{eq:polylog}) and \ref{eq:idpole}) completes the
set which defines four unknown variables $s_1$, $s_2$, $X_1$, $X_2$ at the pole
singularity point.

The free energy is again
\be
f_{\rm neck}=-\alpha \ln s_1 - (1-\alpha) \ln s_2
\ee
It is calculated numerically and the corresponding plots are shown in \fig{fig:6} at
length of the Section \ref{sect:4}.

\end{appendix}

\end{document}